\documentclass[11pt,a4paper]{article}
\usepackage{jcappub}

\usepackage{amstext}
\usepackage{amssymb}
\usepackage{graphicx}
\usepackage{latexsym}
\usepackage{mathrsfs}
\usepackage{calrsfs}
\usepackage{amsfonts}

{\count255=\time\divide\count255 by 60 \xdef\hourmin{\number\count255}
  \multiply\count255 by-60\advance\count255 by\time
  \xdef\hourmin{\hourmin:\ifnum\count255<10 0\fi\the\count255}}

\newcommand{\nn}{\nonumber \\ }

\def\rd{ {\rm d}}
\def\vev#1{ \left\langle #1 \right \rangle }
\def\abs#1{ \left| #1 \right| }

\def\tobs{\Delta}

\begin{document}

\title{Limits on Semiclassical Fluctuations in the Primordial Universe}

\author[a]{Grigor Aslanyan,}
\author[b]{Aneesh V.~Manohar,}
\author[b]{Amit P.S. Yadav}

\affiliation[a]{Department of Physics, University of Auckland, Private Bag 92019, Auckland, New Zealand}
\affiliation[b]{Department of Physics, University of California at San Diego,
  La Jolla, CA 92093\vspace{4pt} }

\abstract{We place limits on semiclassical fluctuations that might be present in the primordial perturbation spectrum. These can arise if some signatures of
pre-inflationary features survive the expansion, or could be created by whatever mechanism ends inflation. We study two possible models for such remnant fluctuations, both of which break the isotropy of CMB on large scales. We first consider a semiclassical fluctuation in one Fourier mode of primordial perturbations. The second scenario we analyze is a semiclassical Gaussian bump somewhere in space. These models are tested with the seven-year WMAP data using a Markov Chain Monte Carlo Bayesian analysis, and we place limits on these fluctuations. The upper bound for the amplitude of a fluctuation in a single Fourier mode is $a\le 10^{-4}$, while for the Gaussian bump $a\le 10^{-3}$. }

\arxivnumber{1301.5641}

\maketitle

\section{Introduction}

The standard model of cosmology assumes a homogeneous and isotropic universe, and this assumption is in good agreement with the observational data from the CMB (cosmic microwave background) radiation and galaxy surveys~\cite{2012arXiv1212.5226H}. The standard $\Lambda$CDM model invokes a period of exponential expansion, and the observed perturbation spectrum arises from quantum fluctuations in the free-field theory vacuum (Bunch-Davies vacuum) of the inflaton field that are amplified by the expansion of the universe~\cite{Baumann:2009ds}. At present, we do not have a well-defined (renormalizable) quantum field theory for inflation which explains not only the period of exponential expansion, but also how it ends. It is important to test how well the $\Lambda$CDM scenario works, so that one can constrain  field theory models for inflation.

Possible deviations from standard cosmology have been studied in the literature~\cite{Land:2005ad,Land:2006bn,Rakic:2007ve,Aslanyan:2011zp,ArmendarizPicon:2010da}.
It is possible that some anomalies on large scales have evolved directly from the primordial perturbations. If inflation does not last long enough to erase the initial transient contributions to the background dynamics, some observable features may be left in the CMB \cite{Chen:2008wn}. Another possibility of modifying the initial power spectrum is by having inflation that does not start in the Bunch-Davies vacuum state (see, e.g. \cite{Danielsson:2002kx,Easther:2005yr}). Multi-field inflationary models may also generate non-isotropic perturbations~\cite{Linde:1996gt}. The authors of \cite{Gordon:2006ag} have discussed the possibility of a linearly modulated primordial power spectrum. They found an improvement in $\chi^2$ of about $9$ for $3$ extra parameters, using the three-year WMAP data. 

In this paper we study the possibility that the primordial density perturbation spectrum, in addition to the standard quantum fluctuations, has a semiclassical contribution. Semiclassical fluctuations arise naturally in field theory if the initial state is not the quantum vacuum, but contains semiclassical field configurations such as topological defects (vortices, monopoles, skyrmions, domain walls, etc.) which can be produced in cosmological phase transitions. We analyze two different models for such fluctuations, and put bounds on their parameters using the seven-year temperature data from the WMAP satellite. The models provide a functional form for the fluctuations with a few parameters, which can be constrained by the data. Firstly, we consider a semiclassical fluctuation in one of the Fourier modes of primordial perturbations. This model could arise if one momentum mode of the inflaton field was not in its quantum ground state. The second model we consider has a Gaussian density bump, and  corresponds to initial conditions where the scalar field deviates from its vacuum value in some region of space. Clearly, we are only sensitive to fluctuations on scales comparable to the size of the observed universe. Much larger wavelength fluctuations would appear as a constant shift in the mean density of our universe.

This paper is organized as follows. In Section \ref{likelihood_sec} we compute the CMB fluctuations, including additional semiclassical contributions. This result  is then implemented in the likelihood calculation for CMB data in Section~\ref{sec:likelihood} for the specific fluctuations we study. The details of the data analysis are given in Section~\ref{data_analysis_sec}.  We present our results in Section~\ref{results_sec}, and we conclude in Section~\ref{conclusions_sec}.

\section{CMB Perturbations with a Semiclassical Contribution}\label{likelihood_sec}

In this section, we describe how to compute the CMB fluctuations including an additional semiclassical contribution. We first discuss the general case of an
arbitrary semiclassical contribution, and then restrict ourselves to the two cases of a single Fourier mode, or a Gaussian bump, which are analyzed in this paper.

We start with a quick review of the standard CMB analysis, and then discuss its modification. The initial perturbations can be expressed in terms of the initial gauge invariant curvature perturbations $\zeta(\mathbf{k})$ on uniform density hypersurfaces. For homogeneous and isotropic Gaussian perturbations the initial curvature perturbations are completely described by the two-point correlation function, the power spectrum
\begin{equation}\label{matter_P1}
\vev {\zeta(\mathbf{k})\zeta^*(\mathbf{k}^\prime ) } \equiv (2\pi)^3\delta^3(\mathbf{k}-\mathbf{k}^\prime)P(k)\,.
\end{equation}
Higher order correlations of $\zeta$ can be given in terms of the two-point correlations if the fluctuations are Gaussian. The power spectrum $P(k)$ is the usual scale invariant Harrison-Zel'dovich spectrum if the inflationary theory is a free scalar field.

The CMB observable is the temperature anisotropy $\Theta(\mathbf{x},\hat{\mathbf{n}})$, the temperature fluctuation in direction $\hat{\mathbf{n}}$ as seen by on observer located at $\mathbf{x}$.\footnote{We only have data when $\mathbf{x}=\mathbf{x}_0$, our location in the universe. We choose coordinates so that $\mathbf{x}_0=0$.} It is more convenient to Fourier transform the $\mathbf{x}$ dependence, and use $\Theta(\mathbf{k},\hat{\mathbf{n}})$.
The temperature anisotropies $\Theta(\mathbf{k},\hat{\mathbf{n}})$ can be expressed in terms of the initial gauge invariant curvature perturbations $\zeta(\mathbf{k})$

\begin{equation}
\Theta(\mathbf{k},\hat{\mathbf{n}})=\zeta(\mathbf{k})\frac{\Theta(k, \mathbf{k}\cdot\hat{\mathbf{n}})}{\zeta(k)}
\end{equation}
where the ratio $\Theta(k, \mathbf{k}\cdot\hat{\mathbf{n}})/\zeta(k)$ does not depend on the initial curvature perturbations. It is determined from the evolution of $\Theta$ and $\zeta$, and only depends on the magnitude of $\mathbf{k}$ and the direction of $\hat{\mathbf{n}}$ relative to $\mathbf{k}$~\cite{dodelson}.
The angular dependence of the temperature anisotropies is written in terms of the spherical harmonics,
\begin{equation}\label{ylm_decomposition1}
\Theta(\mathbf{x},\hat{\mathbf{n}}) = \sum_{lm}a_{lm}(\mathbf{x})Y_{lm}(\hat{\mathbf{n}})
\end{equation}
and 
\begin{equation}\label{a_lm1}
a_{lm}(\mathbf{x})=\int\frac{\rd^3k}{(2\pi)^3}e^{i\mathbf{k}\cdot\mathbf{x}}\int \rd\Omega\ Y_{lm}^*(\hat{\mathbf{n}})\Theta(\mathbf{k},\hat{\mathbf{n}})\,.
\end{equation}

The observed CMB map $\Theta(\mathbf{x}_0,\hat{\mathbf{n}})$ allows one to measure the two-point temperature correlation
\begin{equation}\label{M5}
M_{lml'm'}\equiv \vev{ a_{lm}(\mathbf{0}) a_{l'm'}^*(\mathbf{0}) }\,.
\end{equation}
All higher order temperature correlations are given in terms of this if the fluctuations are Gaussian.

Expanding $\Theta(k, \mathbf{k}\cdot\hat{\mathbf{n}})$ into Legendre polynomials
\begin{equation}\label{theta_expansion1}
\Theta(k, \mathbf{k}\cdot\hat{\mathbf{n}})=\sum_{l}(-i)^l(2l+1)P_l(\hat{k}\cdot\hat{\mathbf{n}})\Theta_l(k)
\end{equation}
the covariance matrix (\ref{M5}) takes the form
\begin{equation}\label{M_inf1}
M_{lml^\prime m^\prime }=\delta_{ll^\prime }\delta_{mm^\prime }C_l
\end{equation}
with
\begin{eqnarray}\label{C_l1}
C_l &=& \frac{2}{\pi}\int \rd k\ k^2\, P(k)\left|R_l(k) \right|^2\,,\nn
R_l(k) &=& \frac{\Theta_l(k)}{\zeta(k)}\,,
\end{eqnarray}
which is the theoretical prediction for the CMB power spectrum, and can be compared with observations.

In our models, the fluctuation $\zeta$ can be written as
\begin{eqnarray}\label{zeta}
\zeta(\mathbf{x}) &=& \zeta_q(\mathbf{x}) + \zeta_{\text{cl}}(\mathbf{x})
\end{eqnarray}
where $\zeta_q$ and $\zeta_{\text{cl}}$ are the  quantum and classical components of $\zeta$. Such a form arises naturally in a quantum field theory; for example, the quantum field around a semiclassical background, such as a soliton, has exactly the same decomposition as in Eq.~(\ref{zeta}), $\phi(\mathbf{x}) =\phi_q(\mathbf{x}) + \phi_{\text{cl}}(\mathbf{x})$, where $\phi_{\text{cl}}(\mathbf{x})$ is the soliton field configuration~\cite{Rajaraman:1982is}. 

The usual assumption for the quantum fluctuations is that the one-point average vanishes,
\begin{eqnarray}
\vev{\zeta_q(\mathbf{x})} &=& 0
\end{eqnarray}
and the two-point correlation is as in Eq.~(\ref{matter_P1}). The additional semiclassical component $\zeta_{\text{cl}}$ is an ordinary function, not a quantum operator, so that
\begin{eqnarray}\label{zeta1}
\vev{\zeta_{\text{cl}}(\mathbf{x})} &=& \zeta_{\text{cl}}(\mathbf{x}) \,,\nn
\vev{\zeta_{\text{cl}}(\mathbf{x})\zeta_q(\mathbf{y})} &=& 0 \,, \nn
\vev{\zeta_{\text{cl}}(\mathbf{x})\zeta_{\text{cl}}(\mathbf{y})} &=& \zeta_{\text{cl}}(\mathbf{x}) \zeta_{\text{cl}}(\mathbf{y}) \,.
\end{eqnarray}
In Fourier space, this gives
\begin{eqnarray}\label{zeta2}
\vev{\zeta(\mathbf{k})} &=& \zeta_{\text{cl}}(\mathbf{k}) \,,\nn
\vev{\zeta(\mathbf{k})\zeta^*(\mathbf{k}^\prime)} &=& \zeta_{\text{cl}}(\mathbf{k}) \zeta^*_{\text{cl}}(\mathbf{k}^\prime)+(2\pi)^3\delta^3(\mathbf{k}-\mathbf{k}^\prime)P(k) \,.
\end{eqnarray}
for the initial $\zeta$ spectrum. 

Combining Eq.~(\ref{zeta}) with the previous equations, one finds that
\begin{eqnarray}\label{2.13}
a^{(\text{cl})}_{lm} \equiv \vev{a_{lm}(\mathbf{0})} &=& \frac{(-1)^l}{2\pi^2} \int \rd k\, \rd \Omega_k\  k^2\, Y^*_{lm}(\hat{\mathbf{k}})\, R_l(k) \, \zeta_{\text{cl}}(\mathbf{k})
\end{eqnarray}
and the two-point correlation is
\begin{eqnarray}
M_{lml^\prime m^\prime}=\vev{ a_{lm}(\mathbf{0}) a_{l'm'}^*(\mathbf{0}) } &=& a^{(\text{cl})}_{lm} a^{(\text{cl})*}_{l^\prime m^\prime}+M^{(q)}_{lml^\prime m^\prime}
\end{eqnarray}
where the first term is the new contribution, and the second term is the standard result Eq.~(\ref{M_inf1},\ref{C_l1}), now denoted by a superscript $q$ to distinguish it from the total $M$. In general, the $n$-point correlations of the shifted quantities $a_{lm}-a_{lm}^{(\text{cl})}$ have the same value as in the theory without a semiclassical fluctuation. This is true even if the correlations are not Gaussian.

To analyze the experimental data, it is also useful to have formul\ae\ for the temperature correlations in pixel space. The temperature $\tobs_i$ in a pixel in direction $\hat{\mathbf{n}}_i$ is
\begin{equation}
\tobs_i=\int \rd\hat{\mathbf{n}}\,\Theta(\hat{\mathbf{n}})B_i(\hat{\mathbf{n}})
\end{equation}
where $B_i(\hat{\mathbf{n}})$ is the beam pattern of pixel $i$ (including the pixel window function).
The beam pattern is specific to the experiment. Usually the beam patterns have the same shape for every pixel and are axially symmetric around the center of the pixel, as is the case for WMAP. If we denote the direction to the center of the pixel by $\hat{\mathbf{n}}_i$ then the beam pattern can be decomposed into spherical harmonics
\begin{equation}\label{B_decomposition1}
B_i(\hat{\mathbf{n}})=\sum_{lm}B_lY_{lm}(\hat{\mathbf{n}}_i)Y_{lm}^*(\hat{\mathbf{n}})\,,
\end{equation}
and $B_l$ does not depend on the pixel.
Then
\begin{eqnarray}\label{2.17}
\tobs_i^{(\text{cl})} &\equiv& \vev{ \tobs_i} = \sum_l (2l+1)(-1)^l \int \frac{\rd^3 \mathbf{k}}{(2\pi)^3}B_l\, P_l(\hat{\mathbf{k}} \cdot \hat{\mathbf{n}}_i)\, R_l(k) \, \zeta_{\text{cl}}(\mathbf{k})\,,
\end{eqnarray}
and the two-point pixel-pixel temperature correlation is
\begin{eqnarray}\label{2.18}
\vev{ \tobs_i \tobs_j} = \tobs_i^{(\text{cl})} \tobs_j^{(\text{cl})}+\vev{\tobs_i \tobs_j}_q
\end{eqnarray}
where $\vev{\tobs_i \tobs_j}_q$ is the standard (quantum) contribution
\begin{eqnarray}\label{2.19}
\vev{\tobs_i\tobs_j}_q &=& \sum_{lml^\prime m^\prime }M^q_{lml^\prime m^\prime }B_lB_{l^\prime }^*Y_{lm}(\hat{\mathbf{n}}_i)Y_{l^\prime m^\prime }^*(\hat{\mathbf{n}}_j)\nn
&=& \sum_{lm} C_l \abs{B_l}^2 Y_{lm}(\hat{\mathbf{n}}_i)Y_{l m }^*(\hat{\mathbf{n}}_j)\,.
\end{eqnarray}
Equations~(\ref{2.17},\ref{2.18},\ref{2.19}) form the basis for the likelihood analysis of the next section.

\section{Likelihood Calculation}\label{sec:likelihood}

We use the likelihood method to compare the theory with the experimental data. The likelihood is calculated in real space. For $N_p$ pixels the likelihood function has the form
\begin{equation}\label{likelihood_def1}
\mathcal{L}=\frac{1}{(2\pi)^{N_p/2}(\det C)^{1/2}}\exp\left(-\frac{1}{2} {\tobs}^T C^{-1}\tobs \right)
\end{equation}
where $\tobs$ is the vector of pixels of measured temperature anisotropies and $C_{ij}$ is the covariance matrix, including noise. 
The theoretical covariance matrix is obtained by transforming $M_{lml^\prime m^\prime }$ into real space, and is given by Eq.~(\ref{2.19}), 
\begin{eqnarray}\label{cov}
C_{ij} = \vev{\tobs_i \tobs_j}_q,
\end{eqnarray}
to which the noise matrix must be added. The likelihood is a function of the cosmological parameters through the dependence of $C$ on the cosmology.

In the modified theory we consider, there is a one-point function, Eq.~(\ref{2.17}), and a modfied two-point function Eq.~(\ref{2.18}). Thus to analyze the data, we can use the modified likelihood function
\begin{equation}\label{likelihood}
\mathcal{L}=\frac{1}{(2\pi)^{N_p/2}(\det C)^{1/2}}\exp\left(-\frac{1}{2} \left[{\tobs}-\tobs^{(\text{cl})} \right]^T C^{-1}\left[{\tobs}-\tobs^{(\text{cl})} \right]\right)\,,
\end{equation}
replacing the observed temperature values by their difference from the expected mean, $\tobs_i \to \tobs_i-\tobs_i^{(\text{cl})}$. The covariance matrix $C$ remains the quantum covariance matrix Eq.~(\ref{cov}) plus instrumental noise. Equation~(\ref{likelihood}) can be used to analyze any semiclassical fluctuation in the primordial universe. In this paper, we restrict ourselves to two interesting functional forms, the single Fourier mode, and the Gaussian bump.

\subsection{Single Fourier Mode}

Consider the case of a semiclassical perturbation in a single momentum mode. We choose
\begin{eqnarray}
\zeta_{\text{cl}} (\mathbf{x}) &=& a_0\cos(\mathbf{k_0}\cdot\mathbf{x}+\alpha)
\end{eqnarray}
which is given in terms of 5 parameters --- the amplitude $a_0$, the wavelength $\lambda=2\pi/k$ and the direction (the 3 parameters in $\mathbf{k_0}$),
and the phase $\alpha$ at Earth. From Eq.~(\ref{2.13}),
\begin{eqnarray}
a^{(\text{cl})}_{lm} &=& 2\pi a_0(-i)^l \, R_l(k_0)\, Y_{lm}^*(\hat{k}_0)\left(e^{i\alpha}+(-1)^le^{-i\alpha}\right) 
\end{eqnarray}
and
\begin{eqnarray}\label{delta_lp_sum}
\tobs^{(\text{cl})}_i &=&\frac{a_0}{2}\sum_l(2l+1)(-i)^lB_lP_l(\hat{\mathbf{n}}_i\cdot\hat{k}_0)\left(e^{i\alpha}+(-1)^le^{-i\alpha}\right)
R_l(k_0)
\end{eqnarray}

\subsection{Gaussian Fluctuation}

The second case studied, the Gaussian fluctuation, has
\begin{eqnarray}
\zeta^{(\text{cl})} (\mathbf{x})&=& a_0e^{-(\mathbf{x}-\mathbf{x}_c)^2/w^2}
\end{eqnarray}
which also has  $5$ new parameters --- the amplitude $a_0$, the radius of the Gaussian $w$, and the position of the center $\mathbf{x}_c$.
The new contributions are given by
\begin{eqnarray}\label{alm_pert}
a^{(\text{cl})}_{lm}(\mathbf{0})
&=&\frac{2}{\pi}a_0w^3\pi^{3/2}(-1)^l\int k^2\rd k\,e^{-k^2w^2/4} R_l(k)\, j_l(kr)Y_{lm}^*(\hat{\mathbf{r}})
\end{eqnarray}
where we have used $\mathbf{r}=\mathbf{x}_c$. 

\begin{eqnarray}\label{delta_gb_sum}
\tobs^{(\text{cl})}_i  &=& \frac{1}{2\sqrt{\pi}}a_0\sum_l(-1)^l(2l+1)B_lP_l(\hat{\mathbf{n}}_i\cdot\hat{\mathbf{r}}) \int k^2\rd k\,e^{-k^2/4}
R_l(k/w) \ j_l(kr/w)\,.
\end{eqnarray}

We use Eq.~(\ref{likelihood}) for the likelihood function in our data analysis. It depends on the cosmological parameters through $C$, and the parameters in the semiclassical fluctuation through $\tobs^{(\text{cl})}$. In principle, one should maximize the likelihood with respect to all the cosmological parameters as well as the five new parameters. We will use a simpler approach. Since we are  interested in the breaking of isotropy on large scales, we will assume that the periodic fluctuation wavelength is of the same order as the distance to the last scattering surface $L_0=14.4\,\text{Gpc}$. This means that our modification will affect only the low-$l$ part of the covariance matrix. Since the standard cosmological parameters are determined from the whole spectrum of $l$, we fix their values at their best-fit values as given by the  seven-year WMAP data~\cite{Komatsu:2010fb}, and only vary the new parameters.\footnote{The values of the cosmological parameters that we use are $100\Omega_bh^2=2.227$, $\Omega_ch^2=0.1116$, $\Omega_{\Lambda}=0.729$, $n_s=0.966$, $\tau=0.085$, $\Delta_R^2(0.002\,\text{Mpc}^{-1})=2.42\times10^{-9}$.}

\section{Data Analysis}\label{data_analysis_sec}

We use the publicly available CAMB code~\cite{Lewis:1999bs} for calculating the standard covariance matrix and $\Theta_l(k)/\zeta(k)$, and a modification of the likelihood code provided by WMAP~\cite{Jarosik:2010iu,Larson:2010gs,Komatsu:2010fb} for calculating the likelihood. Since our modification affects only the low-$l$ part of the spectrum, we use the low-resolution part of the likelihood code. The details of the low-$l$ likelihood calculation for WMAP data can be found in \cite{Hinshaw:2006ia}. Let us summarize the important points. The sky map used for low-$l$ analysis is the ILC (Internal Linear Combination) map. The map is smoothed to $9.183^\circ$ and degraded to $N_{side}=16$ using the HEALPIX software \cite{Gorski:2004by}. Smoothing is done to reduce the noise in high $l$. The map is further masked with the Kp2 mask, leaving $2482$ pixels out of $3072$. Foreground contamination, as well as instrumental noise in the unmasked region should be negligible \cite{Hinshaw:2006ia}. Although it is not recommended to use the ILC map for high-resolution analysis, the smoothed and degraded ILC map is reliable for low-resolution analysis and is the default sky map used in low-resolution likelihood code by WMAP. For comparison, we also do the analysis on the V, W, and Q frequency bands as well, smoothed and degraded as described above.

It is hard to accurately estimate the noise covariance matrix of the smoothed and degraded maps since the noises in different pixels are highly correlated. A $1\,\mu \text{K}$ white noise is added to each pixel of the ILC map, and a corresponding term to the diagonal elements of the covariance matrix, to aid numerical regularization of the matrix inversion \cite{Hinshaw:2006ia, wmap-supplement}. This is small enough to have a negligible effect on the data, but large enough to dominate over the instrumental noise present in the data. For the high resolution V, W, and Q maps the instrumental noise in each pixel can be estimated by $\sigma=\sigma_0/\sqrt{N_{obs}}$ with $\sigma_0=3.137\,\text{mK}$ (V), $6.549\,\text{mK}$ (W), and $2.197\,\text{mK}$ (Q) \cite{wmap-supplement}. $N_{obs}$ is supplied for each pixel along with temperature in WMAP sky maps. In order to estimate the noise in the smoothed and degraded maps we generate $1000$ simulations of noise at high resolution, then smooth and degrade them. The average noise per pixel is estimated in this way to be $0.45\,\mu \text{K}$ for V, $0.54\,\mu \text{K}$ for W, and $0.37\,\mu \text{K}$ for Q band low resolution maps. Therefore, adding a $1\mu \text{K}$ noise to each pixel will still dominate over the existing noise, so it is safe to analyze these maps the same way as the ILC.

The residual galactic uncertainty is removed by constructing a foreground template $\tobs_f$ and marginalizing the likelihood function over it. This is done as follows. An extra parameter $\xi$ is introduced into the likelihood function (\ref{likelihood_def1})
\begin{equation}\label{likelihood_fore}
\mathcal{L}(C, \xi)=\frac{1}{(2\pi)^{N_p/2}(\det C)^{1/2}}\exp\left(-\frac{1}{2}(\tobs-\xi\tobs_f)^T C^{-1}(\tobs-\xi\tobs_f)\right)\,.
\end{equation}
 The likelihood is then marginalized over $\xi$ to get
\begin{equation}\label{likelihood_marginalized}
\mathcal{L}(C)=\frac{1}{(2\pi)^{N_p/2}(\det C)^{1/2}}\sqrt{\frac{2\pi}{\tobs_f^TC^{-1}\tobs_f}}\exp\left(-\frac{1}{2}\left(\tobs^TC^{-1}\tobs-\frac{(\tobs^TC^{-1}\tobs_f)^2}{\tobs_f^TC^{-1}\tobs_f}\right)\right)\,.
\end{equation}
The corresponding formula for Eq.~(\ref{likelihood}) is Eq.~(\ref{likelihood_marginalized}) with $\tobs\to \tobs-\tobs^{(\text{cl})}$.

The foreground template used in the WMAP likelihood code is obtained by subtracting the ILC map from the V band map. Note that the ILC map is constructed by removing all known foreground contaminations, so the foreground marginalization for ILC analysis has a very small effect on the results. However, it is essential for analyzing the V, W, and Q bands.

The monopole and dipole ($l=0$ and $1$ terms) are removed from the original maps. However, after masking the maps, these terms are re-introduced. Thus the likelihood function needs to be marginalized over the monopole and the dipole contributions. This is done by introducing large variance $l=0$ and $1$ terms into the covariance matrix.

\begin{figure}[t]
\centering
\includegraphics[width=8cm]{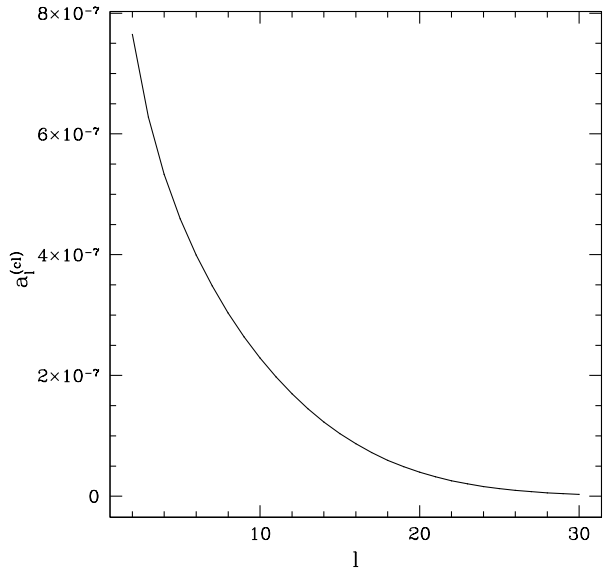}
\caption{\label{unlensed_temp_fig} For a Gaussian fluctuation with $a_0=10^{-3}$, we show $a^{(\text{cl})}_{l}$ in the units of the CMB temperature $2.73\,\text{K}$ as a function of $l$.
}
\end{figure}

Since we are interested in large scale isotropy breaking, we restrict our analysis to $l\le30$. We check that cutting off at $l_{max}=30$ does not have a significant effect on our results for the scales we consider. For the fluctuation in one Fourier mode, the $l=15$ terms in Eq.~(\ref{delta_lp_sum}) are already $3$ orders of magnitude smaller than the $l=2$ terms for the wavelengths $\lambda$ under consideration. This means that the effect of the cutoff can be completely neglected. For the Gaussian bump, the decay with $l$ is not as fast. In Fig.~\ref{unlensed_temp_fig} we plot the dependence of
\begin{equation}\label{alm_avg_eq}
a^{(\text{cl})}_{l}\equiv\frac{1}{2l+1}\sum_{m=-l}^l|a^{(\text{cl})}_{lm}|
\end{equation}
on $l$ for a typical scale $w=0.14L_0=2\,\text{Gpc}$ and a typical amplitude of $a_0=10^{-3}$, keeping all the other parameters fixed. The $l=30$ term is $2$ orders of magnitude smaller than the $l=2$ term, therefore the cutoff cannot have a significant effect in this case either.

The CMB photons get lensed by the matter present in the universe on their journey from the last scattering surface to the observer. However, the WMAP seven-year temperature data by itself is not sensitive to lensing on all scales \cite{Feng:2011jx}. Since we are only considering large scales ($l\le30$), the lensing effects become completely negligible. This is why we do not include lensing in our analysis.

We analyze the data using the Bayesian approach. Usually for a large number of parameters the Bayesian approach is computationally easier than the frequentist approach. One of the main computationally efficient methods for Bayesian analysis is the Markov Chain Monte Carlo (MCMC) method (see, e.g., \cite{mcmc}), which we will use to obtain confidence intervals for our parameters in the Bayesian approach.

\section{Results}\label{results_sec}

For both of the models considered, the data is analyzed using the likelihood function in the space of our new $5$ parameters. It is easier to deal with $-2\ln\mathcal{L}$ instead of the likelihood $\mathcal{L}$ itself. The analyses is sensitive to likelihood ratios only, therefore the overall normalization constant does not matter. Equivalently, any constant can be added to $-2\ln\mathcal{L}$ without changing the likelihood ratios. To make it easier to compare our hypothesis to the standard cosmology we choose to normalize the likelihood function such that
\begin{equation}
-2\ln\mathcal{L}_{st}=0
\end{equation}
where $\mathcal{L}_{st}$ is the likelihood for the standard cosmology.

\subsection{Semiclassical Fluctuation in One Fourier Mode}

We construct a Markov Chain using the Metropolis-Hastings algorithm \cite{mcmc} to calculate the posterior probability distribution functions for our newly introduced parameters. The standard cosmological parameters are kept fixed as before. The prior probability distributions are chosen as follows. The distribution for direction is uniform on the whole sphere. We vary $\lambda$ in the range $[0.5L_0,2.6L_0]$, so we choose the prior distribution to be uniform on that interval and $0$ otherwise. The prior distribution for $\alpha$ is uniform on the whole range $[0,2\pi]$. We choose a Gaussian distribution with $\sigma=10^{-2}$ as the prior distribution for the amplitude. This $\sigma$ is chosen small enough not to allow entry into the nonlinear regime but large enough to keep the prior distribution uniform in the region of interest. As we will see later, the posterior distribution converges to $0$ at scales about two orders of magnitude smaller than this $\sigma$, meaning that the prior distribution can be thought of as uniform with very good approximation. We generate one chain of length $10^7$ and throw away the first $10^4$ elements for burnin. We compare results from a few chains of the same length and burnin time to make sure that they agree well.

\begin{figure}
\centering
\includegraphics[width=5cm]{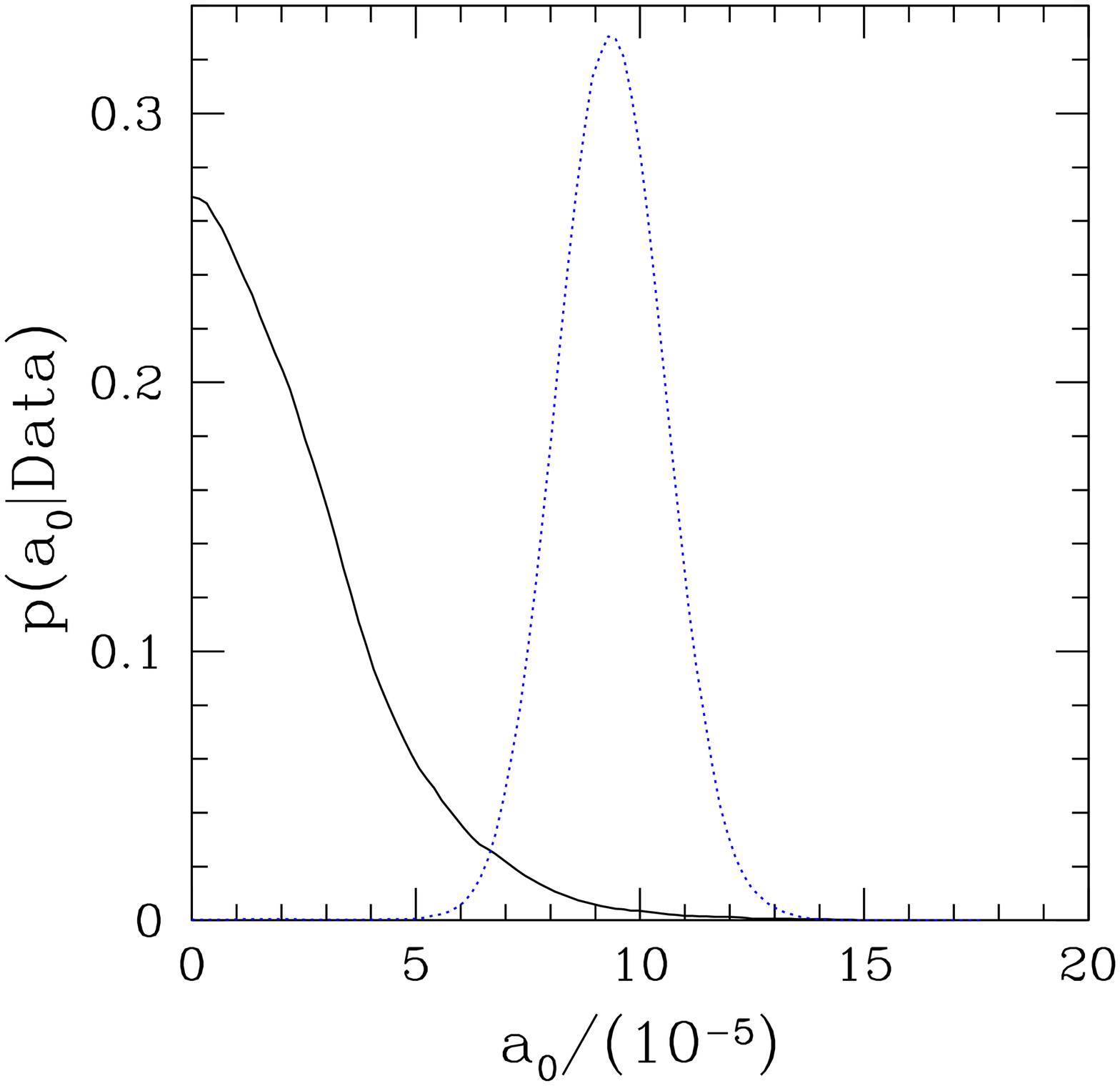}
\includegraphics[width=5cm]{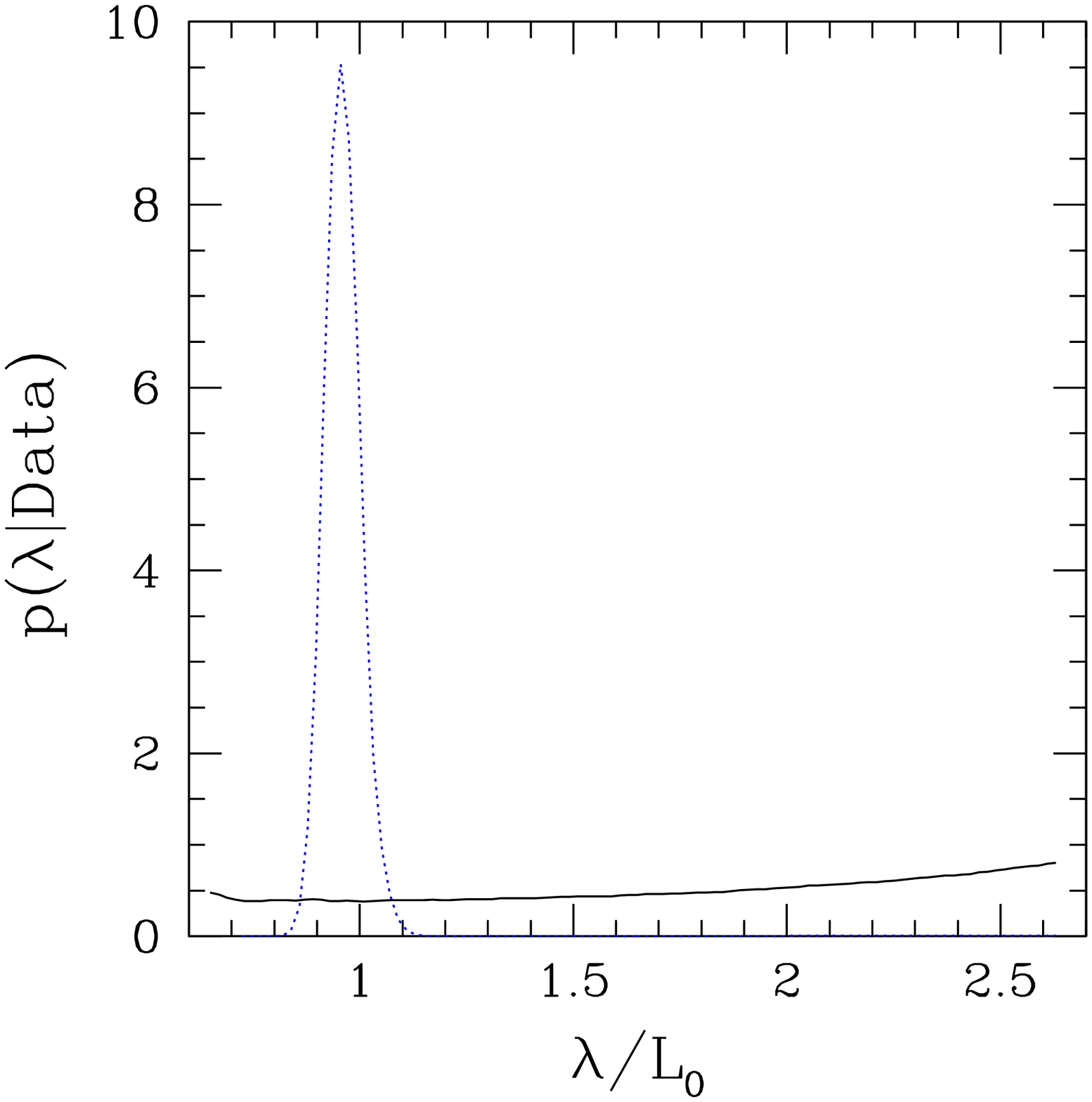}
\includegraphics[width=5cm]{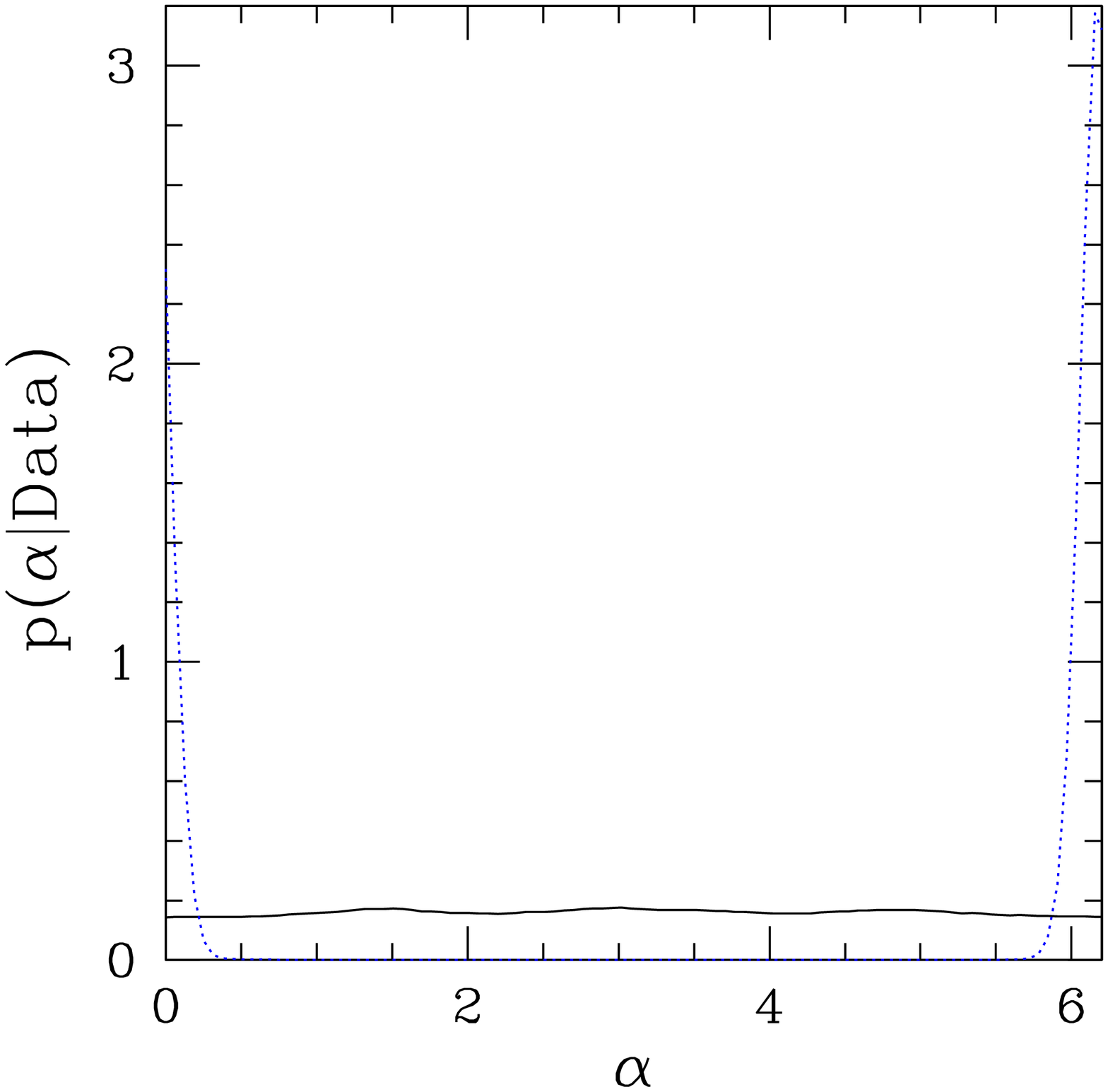}
\caption{\label{sim1} Checks of our pipeline with simulations. We show the posterior distribution $p(x|\text{Data})$ against $a_0$ (left), $\lambda$ (middle), and $\alpha$ (right). In the solid black we show the results from a simulation without the presence of semiclassical fluctuations. In dotted blue curve we show the results from a simulation with an added semiclassical cosine fluctuation with $a_0=10\times 10^{-5}$, $\lambda=L_0$ and $\alpha=0$.}
\end{figure}

To check this method, we first run the analysis on two simulated maps. The first map is a standard universe without any semiclassical fluctuation added. The second simulation has a semiclassical cosine fluctuation added with $a_0=10\times10^{-5}$, $\lambda/L_0=1.0$, $\alpha=0$. The simulations have the same beam function and noise as the real data. The posterior probability distributions for the amplitude $p(a_0|\text{Data})$, the wavelength $p(\lambda|\text{Data})$, and the phase $p(\alpha|\text{Data})$ are shown in Fig.~\ref{sim1}. For simulations without the semiclassical perturbations, the amplitude distribution peaks at $0$ as expected, while the distributions for $\lambda$ and $\alpha$ are essentially flat, without any significant peaks. For simulations with the periodic semiclassical perturbations,  $p(a_0|\text{Data})$ has a clear peak very close to the expected value of $10\times10^{-5}$, $p(\lambda|\text{Data})$ peaks sharply near $1.0\times L_0$, and $p(\alpha|\text{Data})$ has a sharp peak at $\alpha=0$. Those two simulations show that the MCMC method is working very well.

\begin{figure}
\centering
\includegraphics[width=7cm]{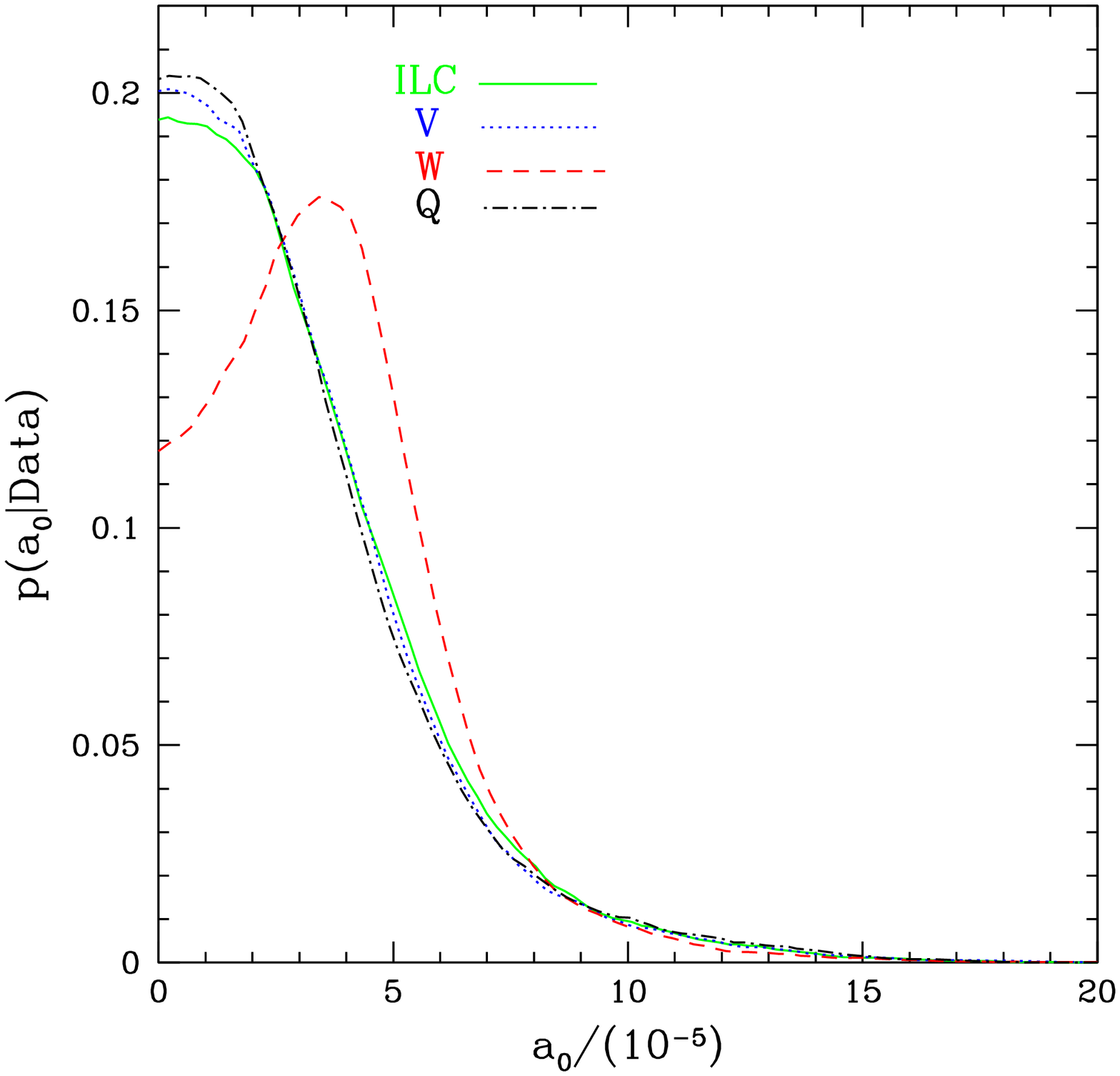}
\includegraphics[width=7cm]{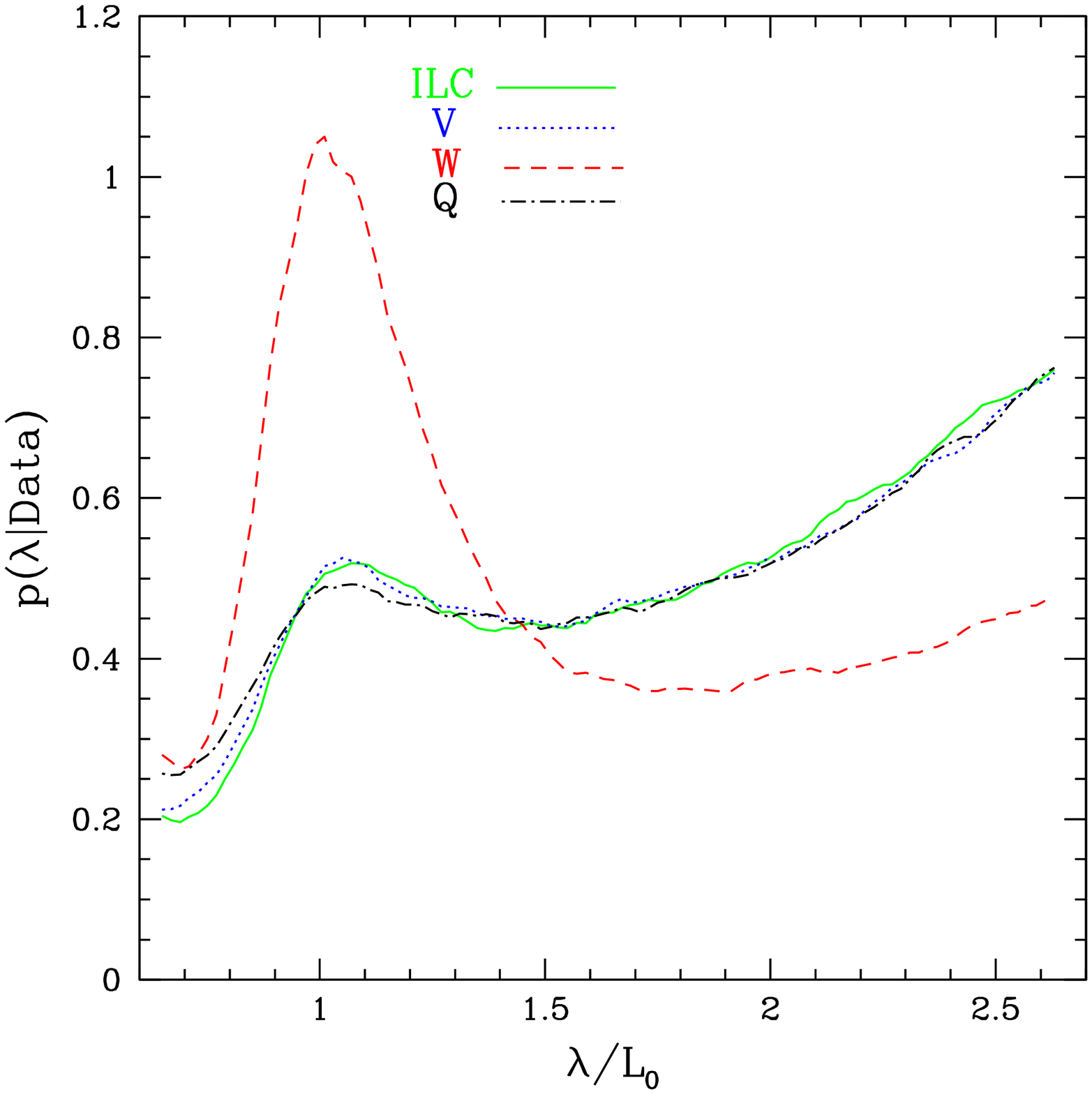}
\includegraphics[width=7cm]{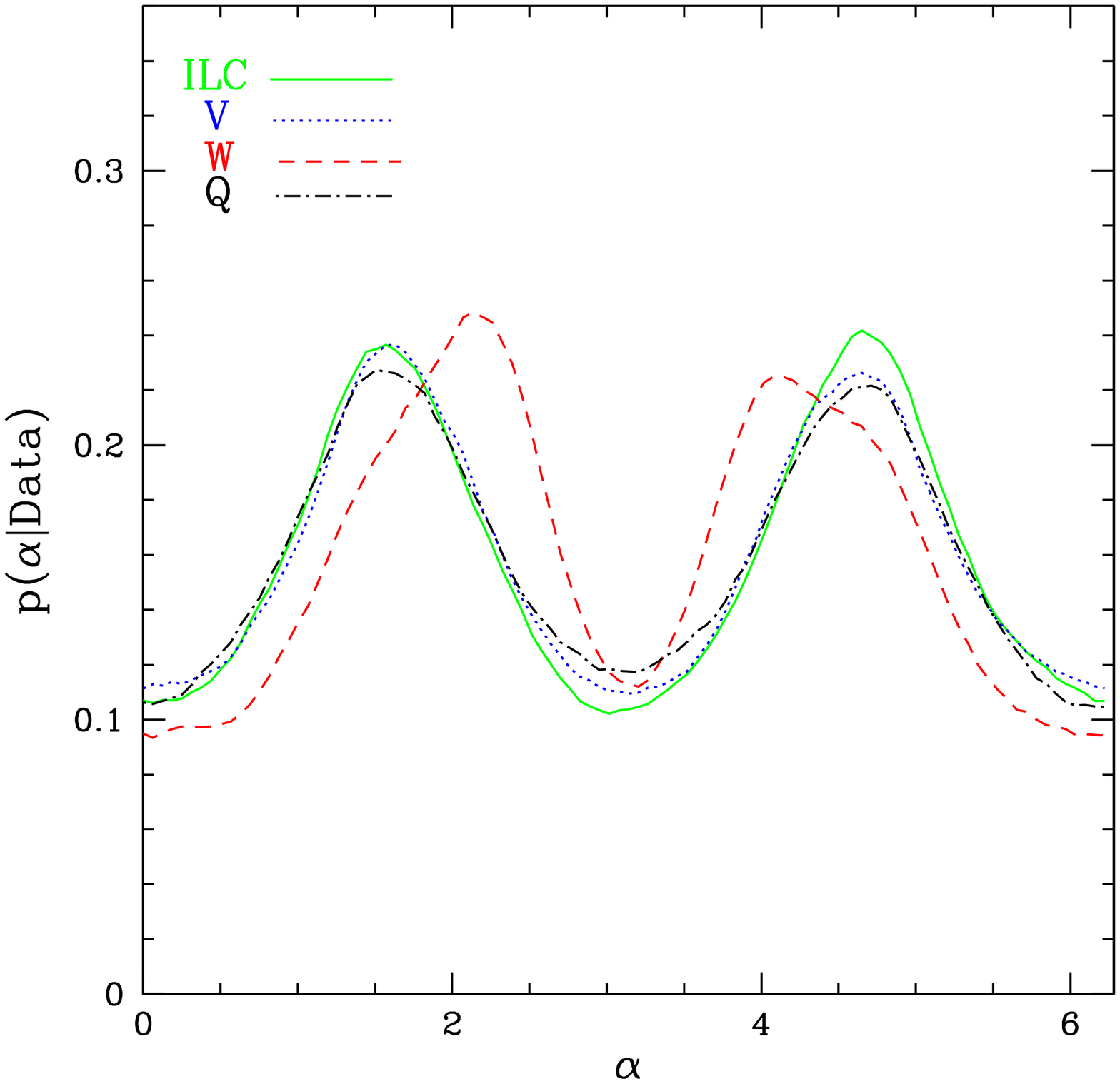}
\raise2cm\hbox{\includegraphics[width=7cm]{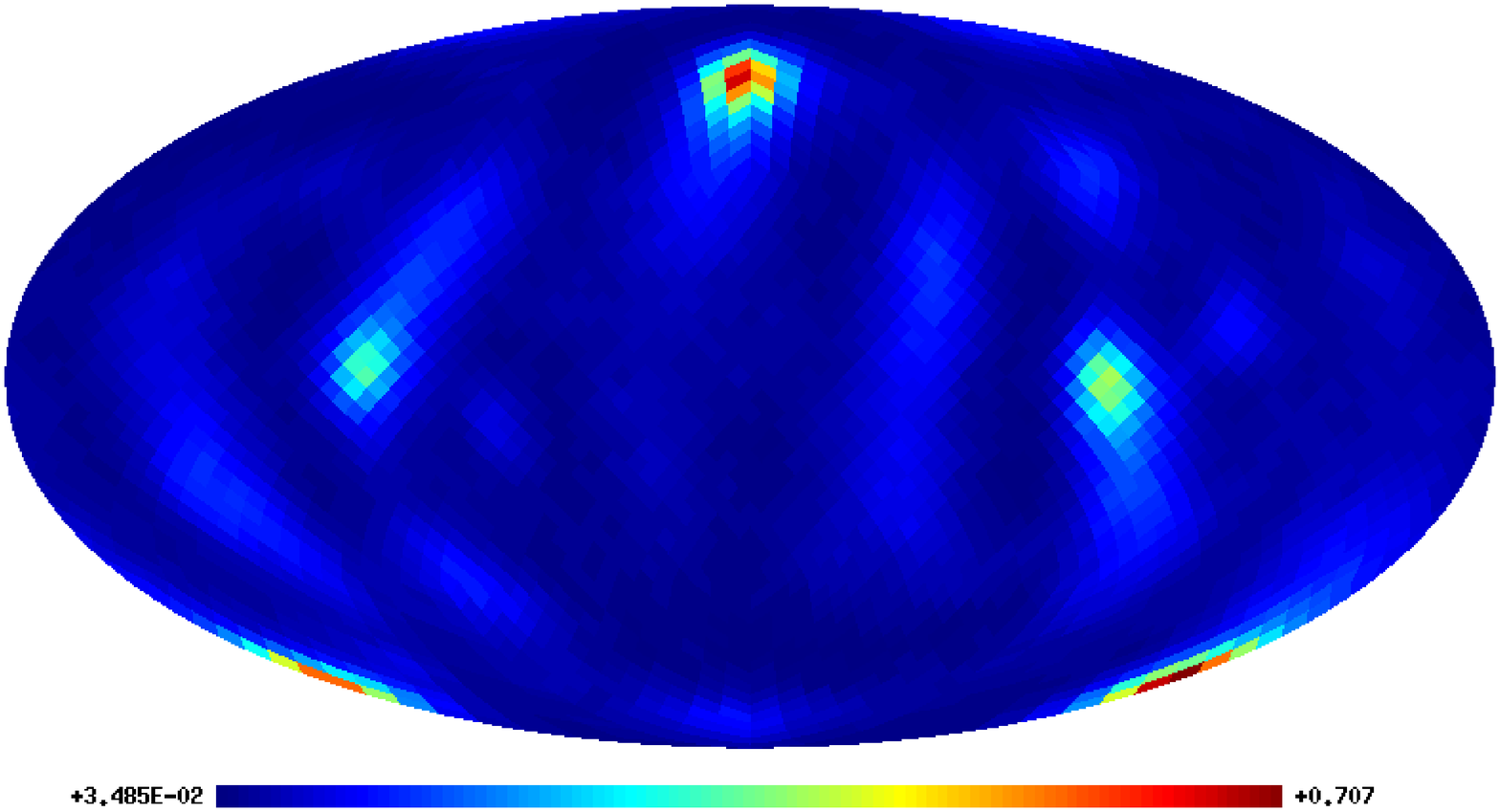}}
\caption{\label{direction_pdf_fig} Results from the WMAP data for a perturbation in a single $k$ mode. We plot the posterior distribution $p(x|\text{Data})$ against $a_0$ (upper left), $\lambda$ (upper right), $\alpha$ (lower left), $\hat{k}^0$ (lower right). In the lower right panel we show the ILC map.
}
\end{figure}

\begin{table}
\renewcommand{\arraystretch}{1.5}
\setlength{\arraycolsep}{5pt}
\begin{eqnarray*}
\begin{array}{c|ccc}
\text{Map} & 68.3 \% & 95.5 \% & 99.7 \% \\
\hline
\text{ILC} & 3.91 & 8.49 & 14.45 \\
\text{V} & 3.81 & 8.42 & 14.84 \\
\text{W} & 4.56 & 8.14 & 14.20 \\
\text{Q} & 3.75 & 8.69 & 14.73 \\
\end{array}
\end{eqnarray*}
\caption{Upper bounds for $|a_0|/(10^{-5})$ for a semiclassical periodic perturbation for different sky maps.} 
\label{amplitude_pdf_limits_cos}
\end{table}

We are now ready to analyze the real sky maps. The posterior distribution functions for the amplitude, the wavelength, the phase, and the direction (ILC map only) are shown in Fig.~\ref{direction_pdf_fig}.  There is no detection for non-zero amplitude for all the maps. The non-zero peak for the W band correspond to less than $2\sigma$. The results from different maps agree reasonably well.  We also obtain upper bounds on $|a_0|$ from $p(a_0|\text{Data})$, which are summarized in Table~\ref{amplitude_pdf_limits_cos}. The limits obtained from different maps agree reasonably well with each other.

\subsection{Semiclassical Gaussian Fluctuation in Space}

\begin{figure}[t]
\centering
\includegraphics[width=7cm]{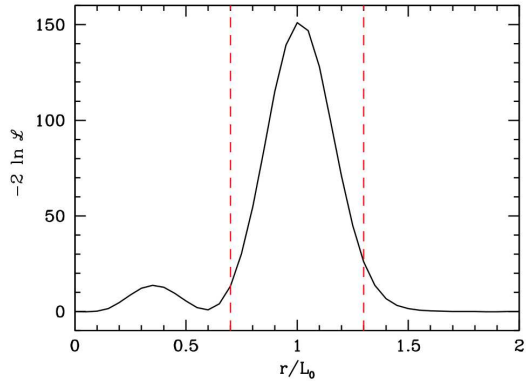}
\caption{\label{r_fixed} Plot of $-2\ln\mathcal{L}$ against $r$ for a Gaussian fluctuation, with all the other parameters fixed ($w=0.3L_0$, $a_0=10^{-3}$). The red dashed lines correspond to $L_0-w$ and $L_0+w$.
}
\end{figure}

We again restrict our analysis to fluctuations on large scales, which means that the parameter $w$ needs to be not much smaller than the distance to the last scattering surface $L_0=14.4\,\text{Gpc}$. We also need to make sure that the fluctuation has a significant causal contact with the last scattering surface, otherwise it will not have an observable effect on the temperature fluctuations. The other issue to keep in mind is that if the center of the bump is very close to our position then the corrections to the temperature fluctuations will be nearly constant and will be absorbed into the constant background temperature. The same thing is true if the center is not very close to us but $w$ is very large. As an example, we plot the dependence of $-2\ln\mathcal{L}$ on $r$ with all the other parameters fixed in Fig.~\ref{r_fixed}. In particular, we choose $w=0.3L_0$ and $a_0=10^{-3}$. As we can see, it is peaked near $L_0$, and the dependence becomes very weak for $r$ outside the range $[L_0-w,L_0+w]$ (red dashed lines). So we will not consider the values of $r$ that are outside that range. This is equivalent to the requirement that the  $1\sigma$ surface of the Gaussian fluctuation must intersect the last scattering surface.

\begin{figure}
\centering
\includegraphics[width=5cm]{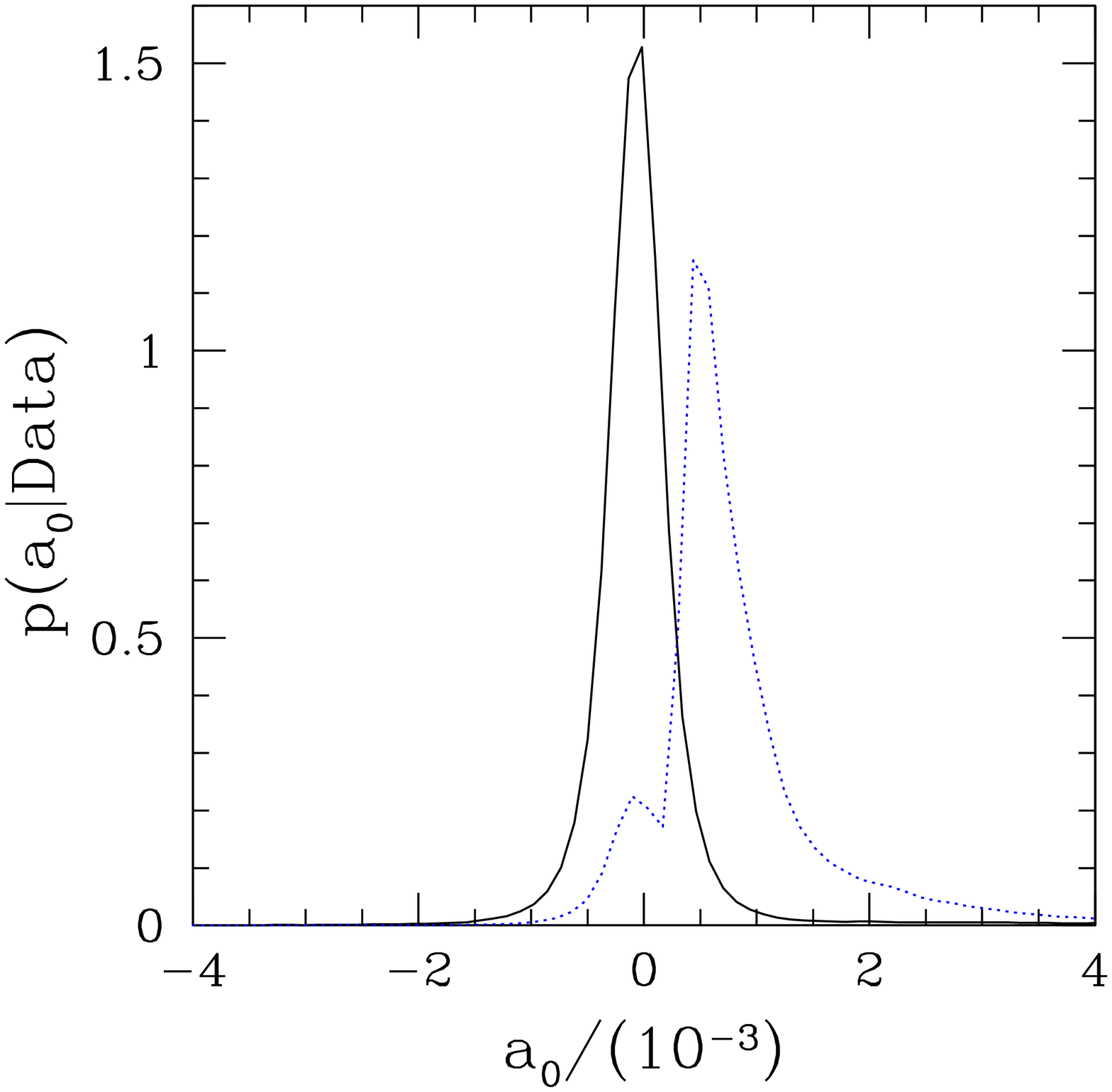}
\includegraphics[width=5cm]{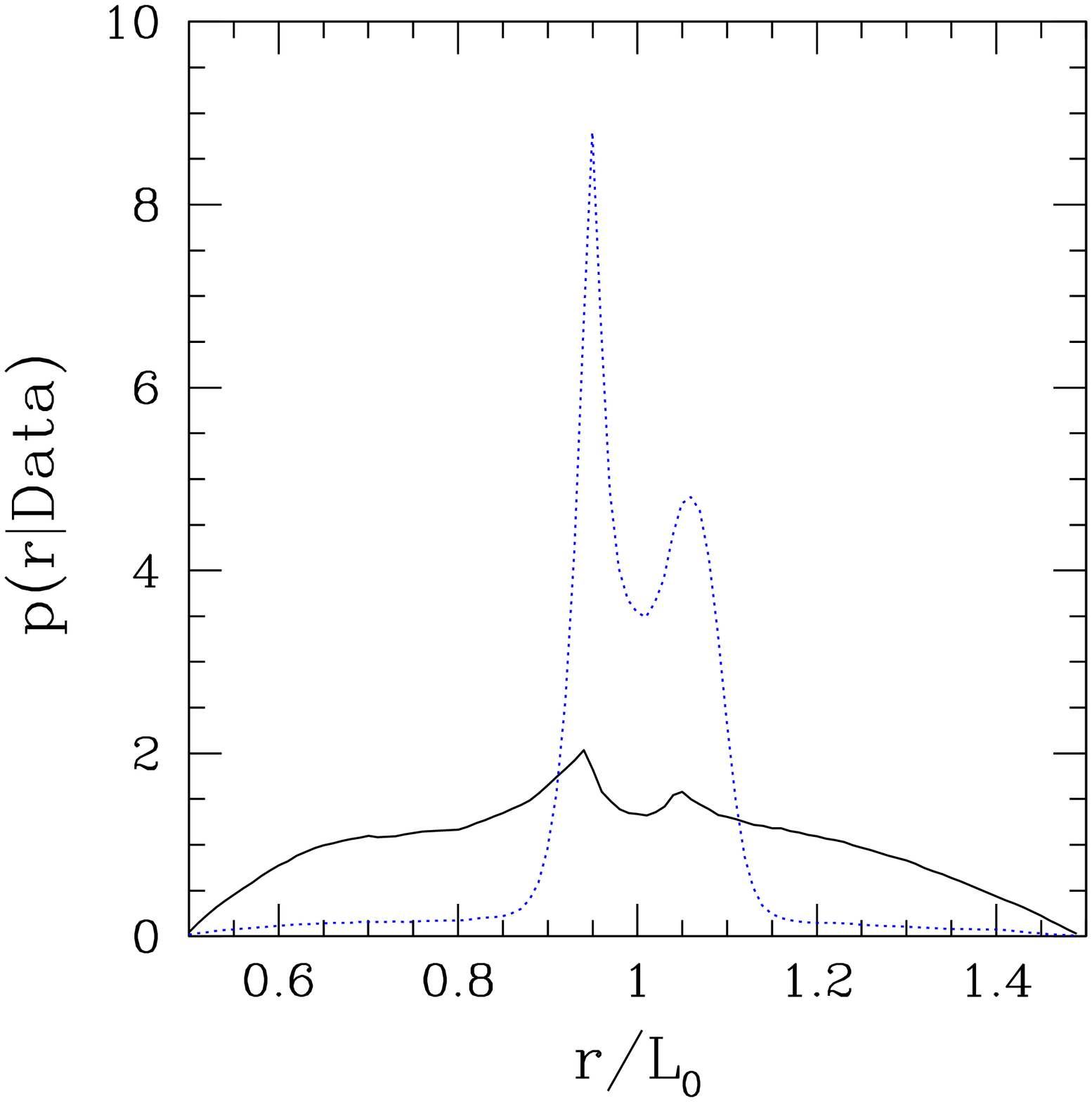}
\includegraphics[width=5cm]{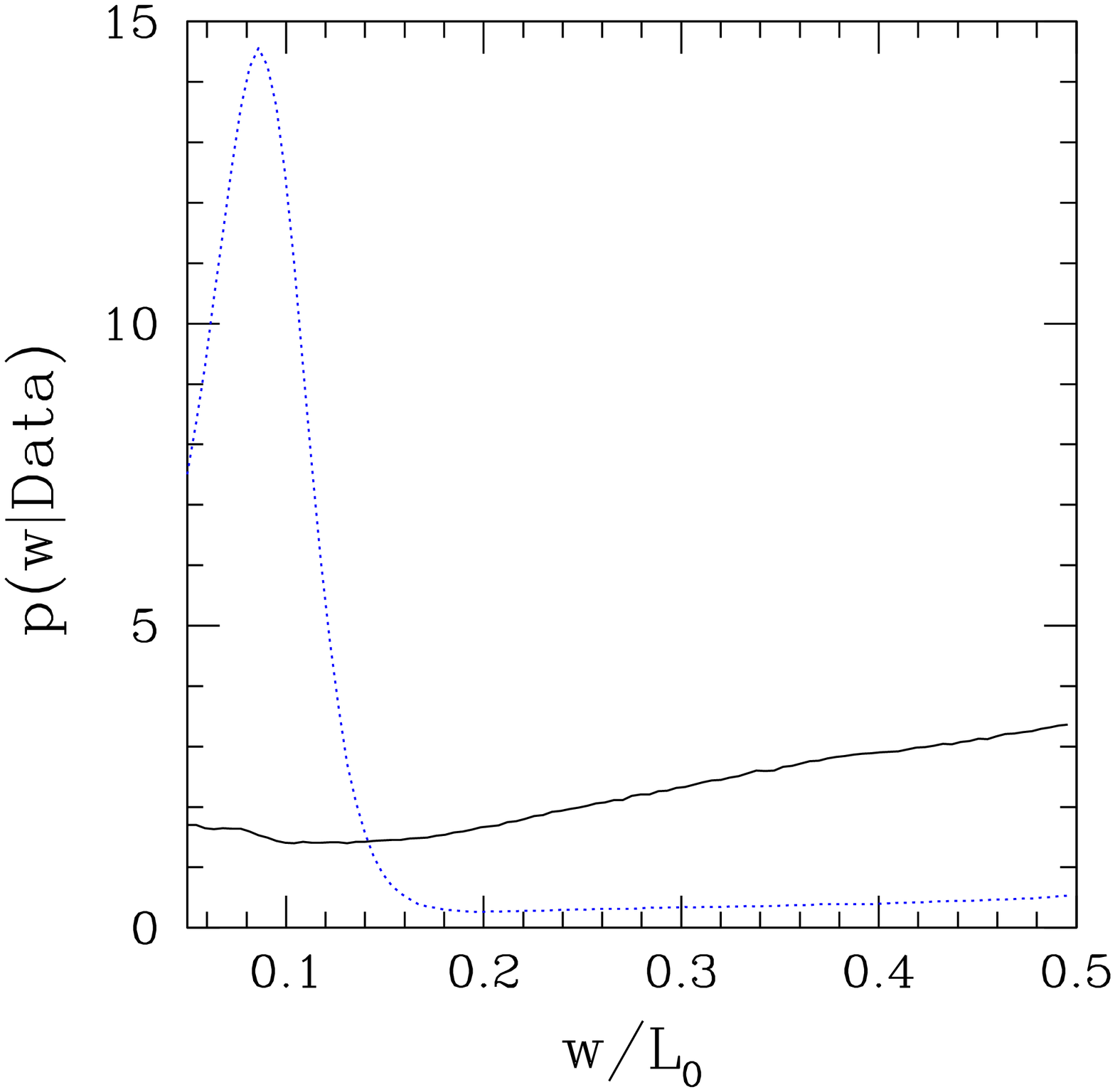}
\caption{\label{sim2} Check of our pipeline for the analysis of a Gaussian bump. We plot the posterior distribution $p(x|\text{Data})$ against $a_0$ (left), $r$ (middle), and $w$ (right). In solid black curve we show the results from a simulation without the presence of semiclassical fluctuations. In the dotted blue curve we show the results from a simulation with a semiclassical Gaussian fluctuation added with $a_0= 0.5 \times 10^{-3}$, $r=0.9\,L_0$ and $w=0.1\,L_0$.
}
\end{figure}

\begin{figure}
\centering
\includegraphics[width=7cm]{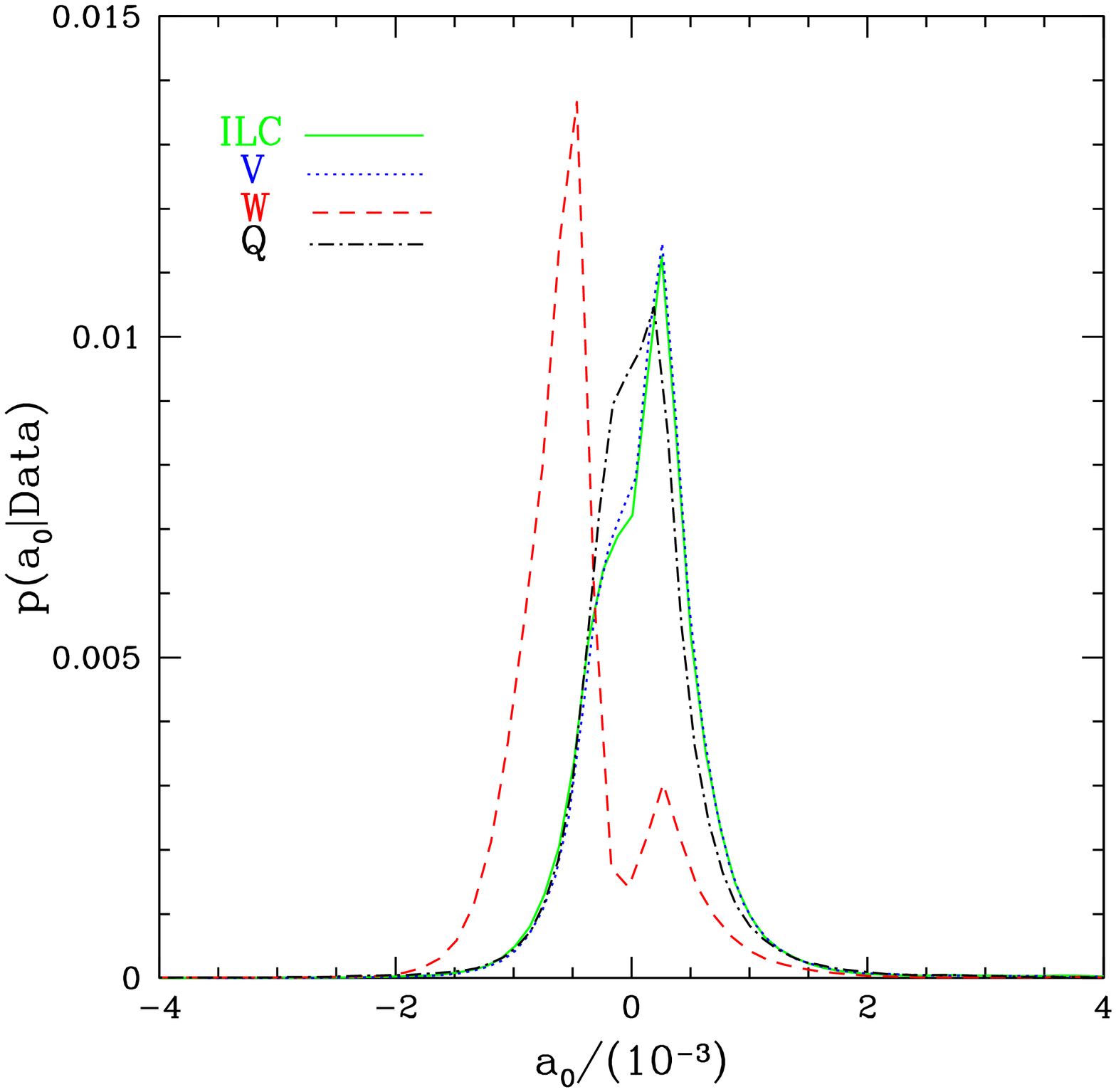}
\includegraphics[width=7cm]{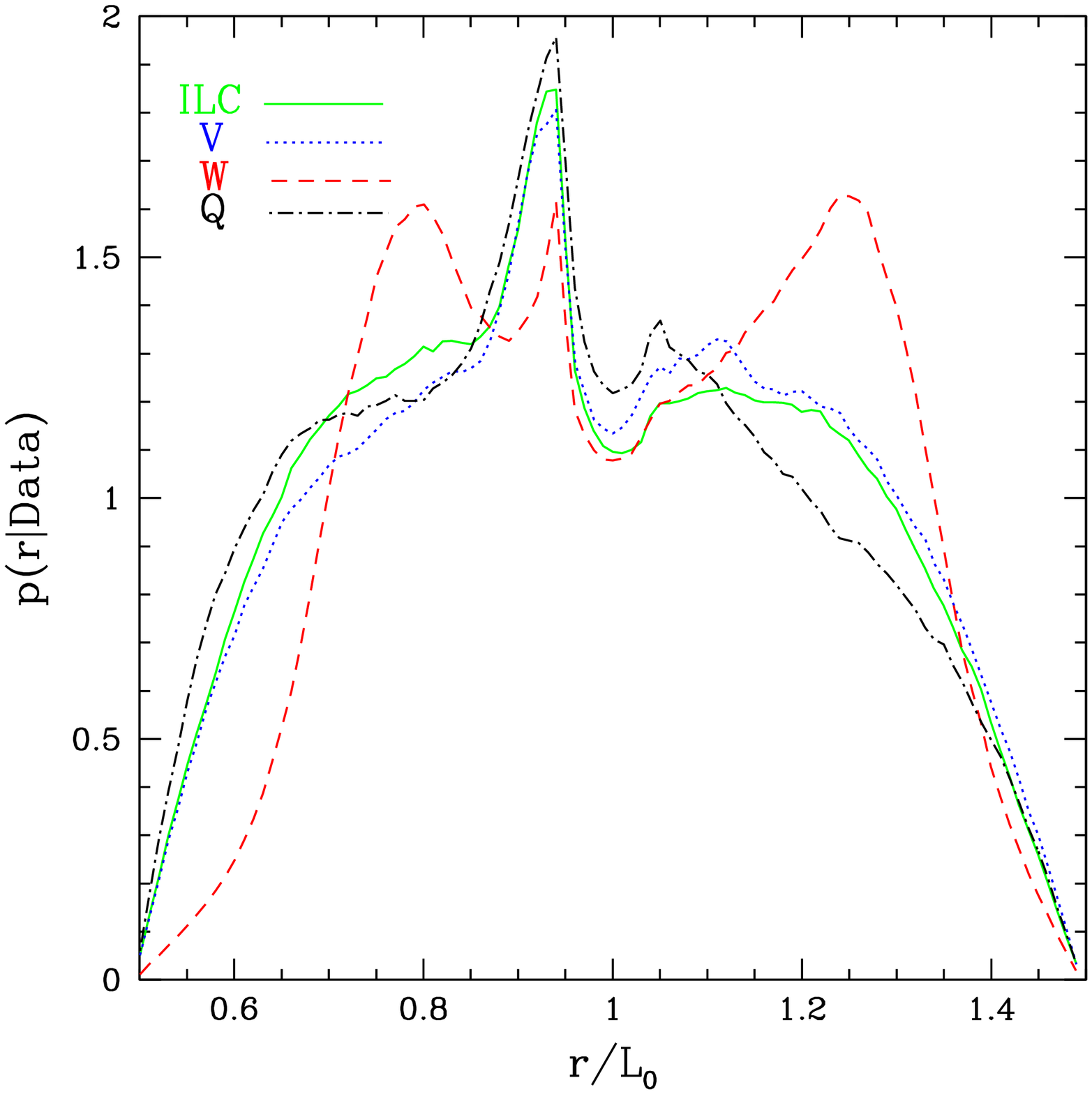}
\includegraphics[width=7cm]{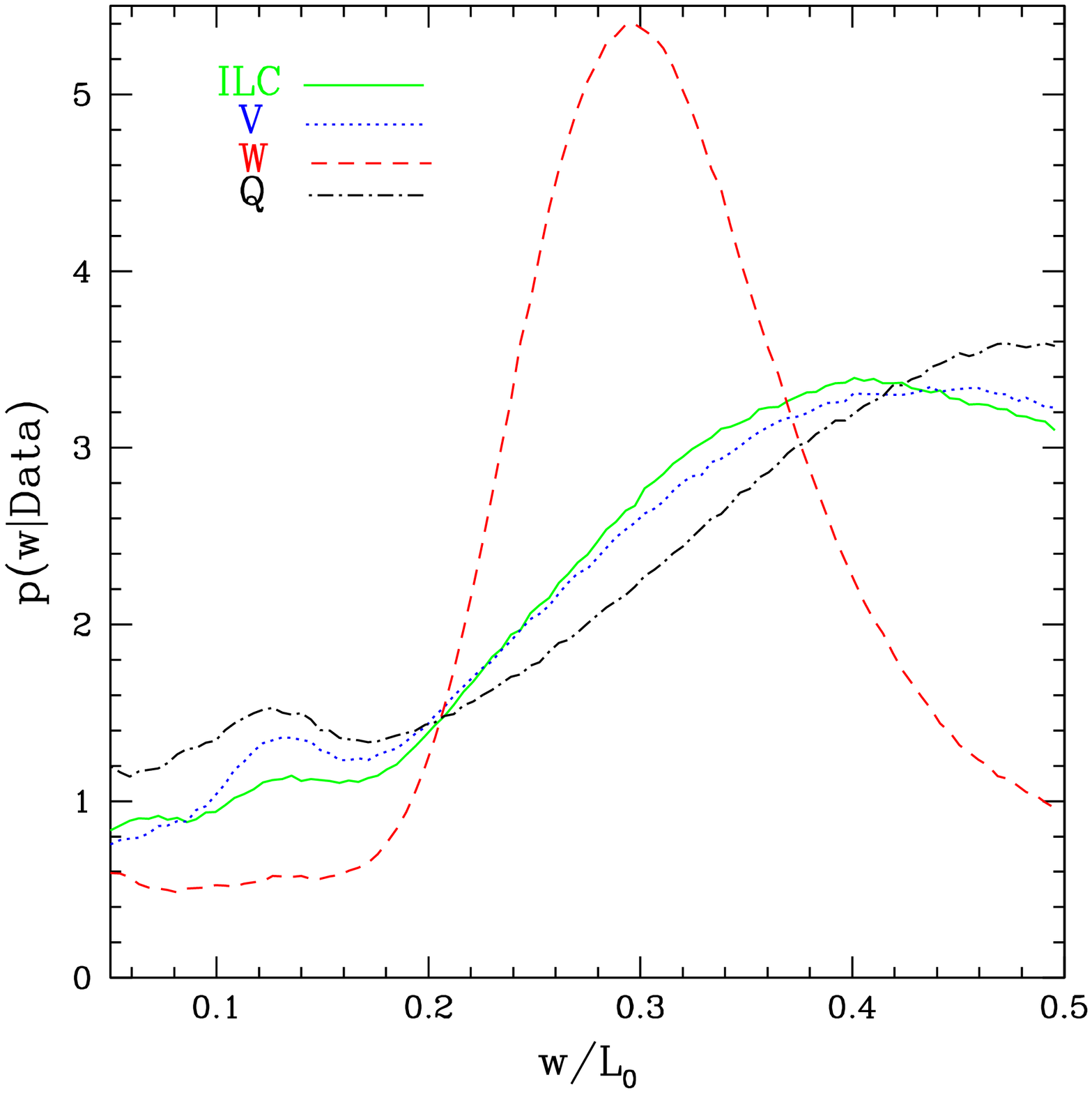}
\raise2cm\hbox{\includegraphics[width=7cm]{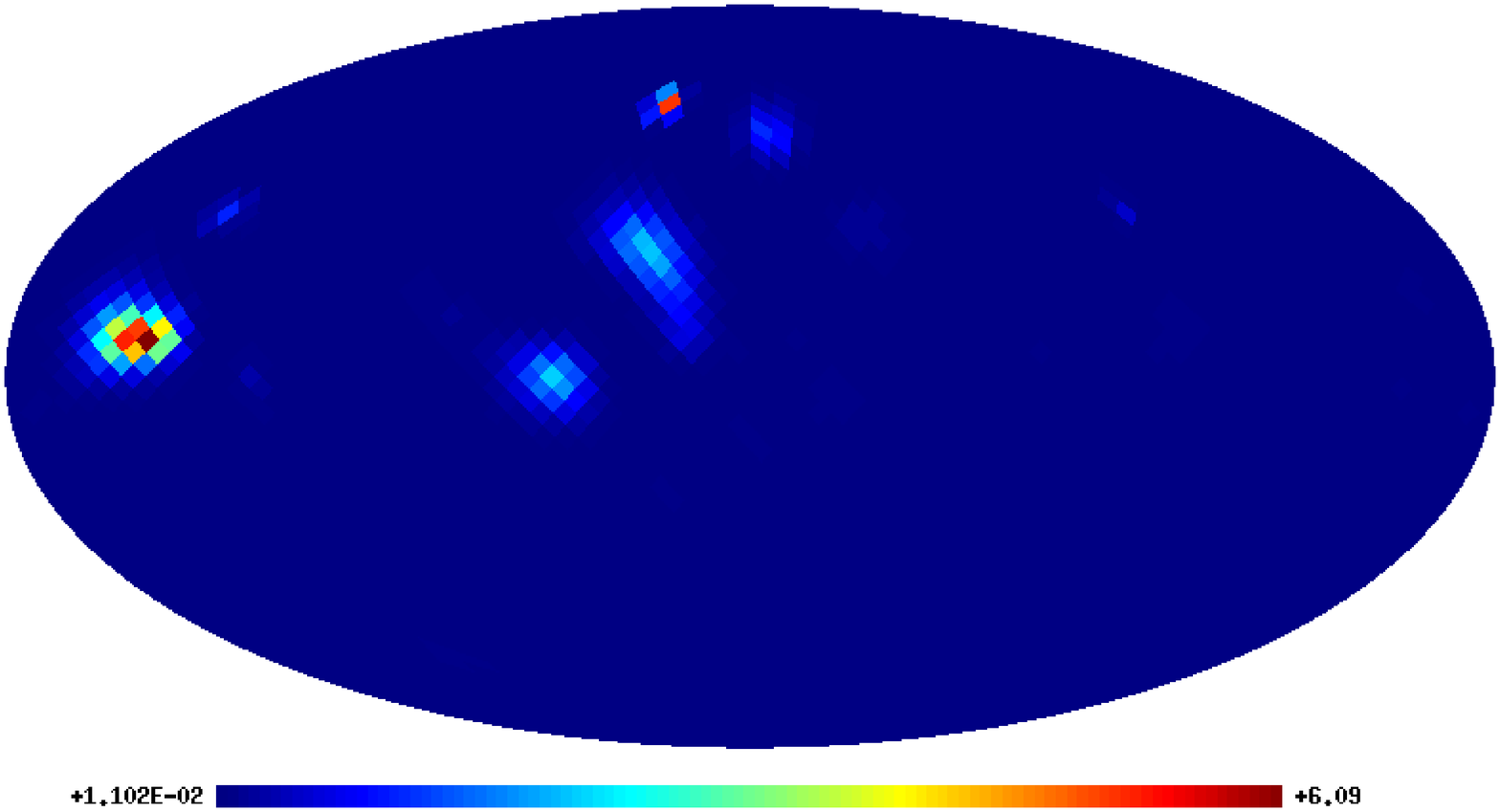}}
\caption{Results from the WMAP data for a Gaussian bump. We plot the posterior distribution $p(x|\text{Data})$ against $a_0$ (upper left), $r$ (upper right), $w$ (lower left), $\hat{r}$ (lower right). In the lower right panel we show the ILC map.
}
\label{WMAP_gaussian} 
\end{figure}

We use the same MCMC algorithm as for the periodic fluctuation to obtain posterior probability distribution functions. For the Gaussian fluctuation the likelihood calculation is computationally slower, so instead of constructing one long chain we construct $10$ chains of length $10^6$, disregarding the first $10^4$ elements of each chain for burnin. We again compare results from a few analyses to make sure that these lengths are sufficient for convergence. The prior distribution for the direction is uniform on the whole sphere. We very $w$ in the range $[0.05L_0,0.5L_0]$ and $r$ in the range $[L_0-w, L_0+w]$ for each value of $w$. The prior distribution for these two parameters is uniform in the area allowed (which is a trapezoid) and $0$ elsewhere. The prior distribution for the amplitude is Gaussian with $\sigma=10^{-1}$. This is again small enough to not allow significant probabilities in the nonlinear regime but large enough to not have a significant impact on the posterior distribution. As we will see, the posterior distribution converges to $0$ about $2$ orders of magnitude earlier, so the prior distribution in the region of interest is essentially uniform.

As for the periodic fluctuation case, we first check the method with simulated maps with and without a Gaussian bump added. Again, the simulated maps have the same beam function and noise as the real maps. The results are shown in Fig.~\ref{sim2}. For a simulation without a Gaussian bump, the distribution for the amplitude $p(a_0|\text{Data})$ peaks at $0$ as expected, while $p(r|\text{Data})$ and $p(w|\text{Data})$ are essentially flat. For the simulations with a Gaussian bump,  the parameters are $a_0=0.5\times10^{-3}$, $r=0.9\, L_0$, $w=0.1\, L_0$. The posterior distributions do indeed have sharp peaks around the expected values. As explained before, for a Gaussian bump the main effect comes from its intersection with the last scattering surface, so we would expect similar results for two bumps of the same size and the same distance from the last scattering surface, but one inside one outside. This explains the smaller peak near $r=1.1\,L_0$. The MCMC method detects the Gaussian bump (or the absence thereof) reliably, and now we can turn to the results from the real data.

\begin{table}
\renewcommand{\arraystretch}{1.5}
\setlength{\arraycolsep}{5pt}
\begin{eqnarray*}
\begin{array}{c|ccc}
\text{Map} & 68.3 \% & 95.5 \% & 99.7 \% \\
\hline
\text{ILC} & 0.46 & 1.15 & 4.95 \\
\text{V} & 0.45 & 1.10 & 4.24 \\
\text{W} & 0.68 & 1.27 & 5.69 \\
\text{Q} & 0.41 & 1.14 & 3.64 \\
\end{array}
\end{eqnarray*}
\caption{Upper bounds for $|a_0|/(10^{-3})$ for a semiclassical Gaussian perturbation for different sky maps.}
\label{amplitude_pdf_limits_gauss}
\end{table}

The results from the analysis of the WMAP data are shown in Fig~\ref{WMAP_gaussian}. The results from all of the maps agree very well with each other except for the W band. For the ILC, V, and Q maps the peaks of the probability distributions lie near the following values of the parameters: $a_0=2.5\times10^{-4}$, $r=0.94\,L_0$, $w=0.15\,L_0$, $b=57^\circ$, $l=29^\circ$. The amplitude probability distribution functions indicate that the amplitude is greater than 0 at about $65\%$ confidence level. For the W map the probability distributions peak at the following values of the parameters: $a_0=-4.6\times10^{-4}$, $r=0.94\,L_0$, $w=0.3\,L_0$, $b=2.4^\circ$, $l=45.0^\circ$. The probability that the amplitude is less than $0$ for the W map is $80\%$. As we can see, for all of the maps the indication of a non-zero amplitude is at a level less than $2\sigma$. The upper bounds for the magnitude of $a_0$ obtained from the probability distribution function are summarized in Table \ref{amplitude_pdf_limits_gauss}.

\section{Discussion and Conclusions}\label{conclusions_sec}
We studied the possibility that the inflationary induced primordial perturbations have a semiclassical contribution in addition to the standard quantum fluctuations. We gave a general formalism for constraining such a primordial semiclassical fluctuation. Then we focused on two specific models of semiclassical perturbations, both of which statistically break CMB isotropy on large scales. The MCMC method that we used works very well as seen by the results from simulated maps with and without semiclassical fluctuations. Although there are some indications of a non-zero amplitude for these fluctuations, they correspond to a confidence level less than $2\sigma$, so our analysis finds no statistically significant evidence for these scenarios. 
We obtained upper bounds on the amplitude of these semiclassical fluctuations at $1$, $2$, and $3$ sigma levels. The bounds for the periodic fluctuation are of the order of $10^{-4}$, and the bounds for the Gaussian bump are of the order of $10^{-3}$. The reason for  weaker bounds in the Gaussian case is that its effect on the CMB strongly depends on the position and the size of the bump. For a bump too far away, the effect on the CMB is very small, while for a bump very close to us, the effect is mostly uniform on the last scattering surface and is subtracted away with the monopole term before the data analysis. Even for a bump with a significant intersection with the last scattering surface, there is a large contribution into the dipole term which is, again, subtracted away before the data analysis.

We used the ILC, V, W, and Q maps from the WMAP seven-year release for our analysis. The results from the different bands, in particular the upper bounds on the amplitude of the semiclassical perturbations, agree reasonably well with each other.

\acknowledgments

The authors would like to thank Richard Easther for helpful discussions and comments on the manuscript, the LAMBDA team for making the seven-year WMAP data and the likelihood calculation software available online, and Antony Lewis and Anthony Challinor for making the CAMB software available online.

The authors wish to acknowledge the contribution of the NeSI high-performance computing facilities and the staff at the Centre for eResearch at the University of Auckland. New Zealand's national facilities are provided by the New Zealand eScience Infrastructure (NeSI) and funded jointly by NeSI's collaborator institutions and through the Ministry of Business, Innovation and Employment's Infrastructure programme. URL {\tt http:/\!/www.nesi.org.nz}. This work was also supported in part by the Department of Energy through DOE Grant No. DE-FG02-90ER40546.

\bibliography{citations}

\end{document}